\documentclass[aps,twocolumn,prl]{revtex4}

\input epsf
\def\plotone#1{\centering \leavevmode
\epsfxsize= 1.0\columnwidth \epsfbox{#1}}

\def\apjl{Astrophys. J. Lett.}
\def\mnras{Mon.Not.Roy.As.Soc.}

\def\aj{Astron. J.}

\def\aap{Astron. \& Astrophys.}
\def\la{\mathrel{\mathpalette\fun <}}
\def\ga{\mathrel{\mathpalette\fun >}}
\def\fun#1#2{\lower3.6pt\vbox{\baselineskip0pt\lineskip.9pt
  \ialign{$\mathsurround=0pt#1\hfil##\hfil$\crcr#2\crcr\sim\crcr}}}

\def\cldd{\C_l^{dd}}
\def\C{{\cal C}}
\def\cleeu{{\tilde \C}_l^{EE}}
\def\clteu{{\tilde \C}_l^{TE}}
\def\clttu{{\tilde \C}_l^{TT}}

\newenvironment{tablehere}{\def\@captype{table}}{}
\newcommand{\tableskip}{\\[-6pt]}

\begin{document}
\bibliographystyle{prsty}
\title{Applications of High Resolution High Sensitivity Observations
  of the CMB}
\author{Manoj\ Kaplinghat}
\affiliation{Department of Physics, One Shields Avenue\\
University of California, Davis, California 95616, USA}
\date{\today}

\begin{abstract}
With WMAP putting the phenomenological standard model of cosmology on
a strong footing, one can look forward to mining the cosmic microwave
background (CMB) for fundamental physics with higher sensitivity and
on smaller scales. 
Future CMB observations have the potential to measure absolute
neutrino masses, test for cosmic acceleration independent of
supernova Ia observations, probe for the presence of dark energy at 
$z \ga 2$, illuminate the end of the dark ages, measure the 
scale--dependence of the primordial power spectrum and detect 
gravitational waves generated by inflation.
\end{abstract}
\maketitle

{\parindent0pt \bf Introduction.} 
The WMAP experiment conclusively showed that the standard cosmological
model is a good phenomenological description of the observed universe
\cite{bennet03a}. In combination with other observations (supernova Ia 
and large scale structure), the consistent picture that emerges has
dark matter contributing about 30\% and dark energy about
70\% to the energy density of the universe. Curvature could contribute
a few \% to the above energy budget but for the rest of this article
we will assume a flat universe.  

Given that the basic phenomenological structure is in place, one can
look forward to the future with some confidence. The CMB has much more
to offer, and in many ways far more spectacular discoveries are waiting
to happen. The aim here will be to lay out a list of things that are
possible (with only a brief description of each topic) with future
observations of the CMB alone. 
Many of the  items on this list are based on
\cite{kaplinghat03c}. Whether some or all of the possibilities on this
list come to fruition depends to large extent on what kind of
polarized foregrounds we will have to  deal with, and if another
full-sky CMB mission after Planck will be funded.   

We will assume that foregrounds will be tame (enough) and that it is a
foregone conclusion another full-sky mission will come to pass.  

{\parindent0pt \bf Damping Tail and Lensing.} 
Recombination is not instantaneous (takes about 1\% of the age of the
universe at the time of recombination) and hence, as recombination
proceeds, photons are able to random-walk and erase anisotropy. The
damping length for this erasure is set by the width of the recombination
surface which is ${\cal O}({\rm Mpc})$. The end result is that the
primary CMB signal (i.e., that produced at the recombination surface)
is damped for multipoles larger than about 1000 (or angles smaller
than about 10 arcminutes).    

Two interesting phenomenon affect the photons after they leave the
recombination surface. First, they are deflected around by all the
inhomogeneities (structure) along their geodesic. Second, the universe
re-ionizes providing energetic electrons for the CMB photons to
interact with. Both of these phenomenon leave imprints on the CMB
which inform us about fundamental physics. 

The dominant signatures from reionization are on large scales
corresponding to the horizon when the universe was reionized. The
lensing effect is significant at much smaller scales. A typical lensing
deflection is ${\cal O}({\rm arcminutes})$. The deflections themselves
are correlated on much larger scales reflecting the correlations in the
underlying mass distribution which causes the lensing. The correlation
peaks on $\sim 10$ degree scales (which makes lensing sensitive to dark
energy clumping).

\begin{figure}[htbp]
  \begin{center}
    \plotone{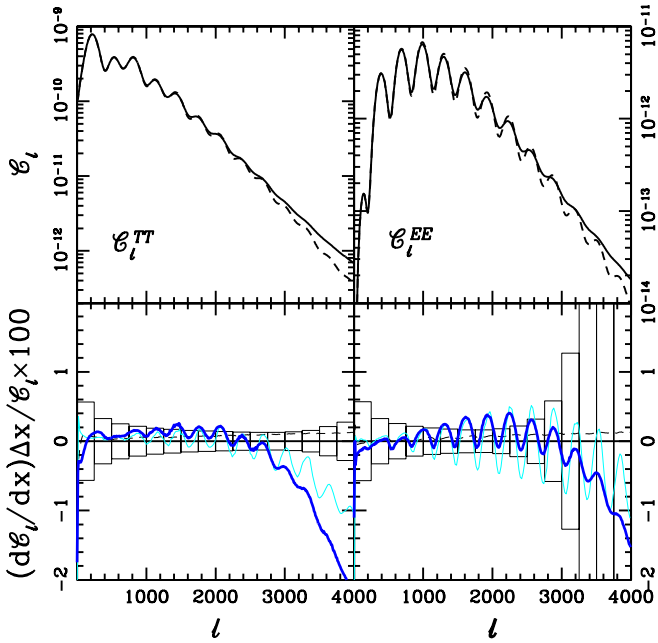}
    \caption{Top panels:  angular power spectra $\C_l^{TT}$ (left)
and $\C_l^{EE}$ (right) for the fiducial model ($m_\nu = 0$) with
no lensing (dashed) and with lensing (solid).  Bottom panels:  
$100 \times d \C_l^{TT} / d m_\nu \times (\Delta m_\nu/\C_l^{TT})$
(dark) and 
$100 \times d \C_l^{TT} / d w_X \times (\Delta w_X/\C_l^{TT})$ (light)
for  $\Delta m_\nu = 0.1$ eV and  $\Delta w_X = 0.2 $. 
Same for $\C_l^{EE}$.
\label{fig:TE}}
\end{center}
\end{figure}

The fact that damping reduces primary CMB power at $l \ga 1000$
is very important. The primary effect of lensing \cite{zaldarriaga98b}
for T and E two-point functions is to average the power over wide
$\ell$ bands, so that at small enough scales most of the power in the
measured TT, EE and ET two-point functions would be due to
lensing (see Fig. \ref{fig:TE}). For polarization, an additional
effect is that lensing converts some of E mode polarization to B
modes.     

\begin{figure}[htbp]
  \begin{center}
    \plotone{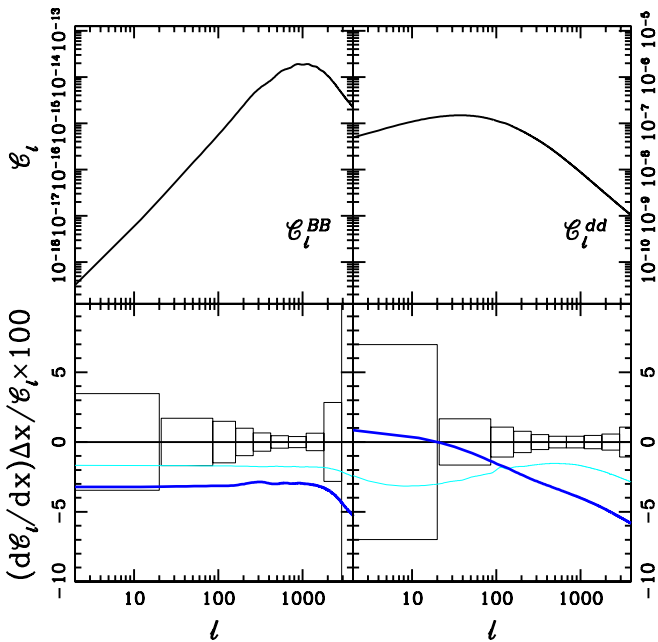}
    \caption{Top panels:  angular power spectra $\C_l^{BB}$ with no
      tensor contribution (left) and $\cldd$ (right) for the fiducial
      model ($m_\nu = 0$). Bottom panels: same as for Fig. 1 but for
      $C_l^{BB}$ and $\cldd$.  
\label{fig:cldd}}
\end{center}
\end{figure}

Lensing introduces calculable non-gaussianities into the CMB
temperature maps
\cite{bernardeau97,zaldarriaga00a,okamoto02,cooray03}. Lets denote the
fourier transform on the sky of measured CMB maps by $a^X(\vec l)$,
$X=\{T,E,B\}$, the corresponding unlensed (unobserved) maps by
$\tilde a^{X}(\vec l)$, and their two point functions by 
${\cal C}_l^{XX'}\equiv 
\langle a^X(\vec l)a^{X'}(\vec l)\rangle l(l+1)/2\pi$ and
$\tilde {\cal C}_l^{XX'}\equiv 
\langle \tilde a^X(\vec l) \tilde a^{X'}(\vec l)\rangle l(l+1)/2\pi$. 
In real space lensing simply shifts a photon
from one part of the sky to another by a small angle (typically $\sim$
arcminutes). In fourier transform space that implies 
$\langle a^X(\vec l)a^{X'}(\vec l') \rangle =   
f_{XX'}(\vec l, \vec l') d(\vec l + \vec l')$
which can be used to estimate the deflection angle $d(\vec L)$
\cite{hu02b} given an  estimate of the unlensed two-point
functions. The power spectrum of the deflection angle is plotted in
the top panel of Fig. \ref{fig:cldd} along with the power spectrum of
the scalar B-mode polarization (created by lensing). 

{\parindent0pt \bf Lensing and Absolute Neutrino Masses.} 
Without the effect of lensing, Eisenstein et al. \cite{eisenstein99}
found that the Planck satellite can measure neutrino mass with an
error of 0.26 eV. This sensitivity limit is related to the temperature
of the recombination surface $T_{\rm rec}\simeq 0.3$ eV.  Neutrinos
with $m_\nu \la T_{\rm dec}$ do not leave imprints on the
recombination surface that would distinguish them from $m_\nu=0$.     

\newcommand{\jeans}{\lambda_J}

Neutrinos with mass $m_\nu \la T_{\rm rec}$ would, however, affect the
evolution of the inhomogeneities. To understand the effects we need to
study the comoving Jeans length for the neutrinos $\jeans (a)$. This
length scale is set by the competition between the pressure due to the 
velocity dispersion of neutrinos and gravity. For non-relativistic
neutrinos in a matter dominated universe  
$\jeans (a) \propto m_\nu^{-1} a^{-1/2}$ \cite{bond83,ma96,hu98}. 
Note that the comoving Jeans length was much larger in the
past. For $k \la k_J(a) \equiv 1/\jeans (a)$, neutrinos can cluster
like dark matter.  On smaller scales, neutrinos do not fall into dark
matter potential wells.   

\begin{figure}[htbp]
  \begin{center}
    \plotone{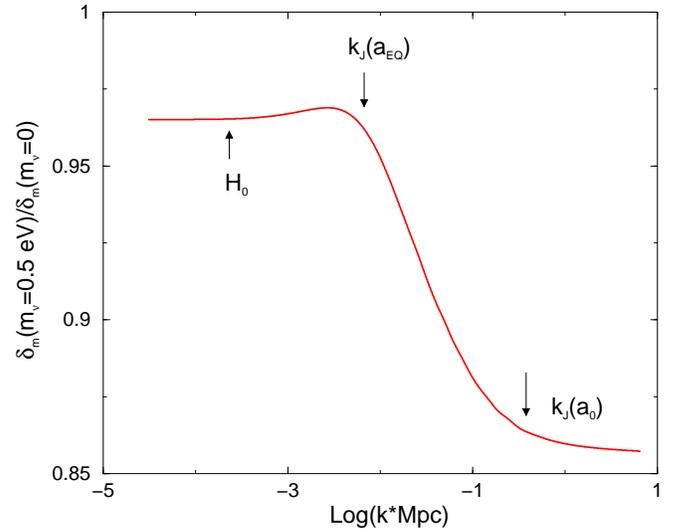}
    \caption{Effect of a massive neutrino on dark matter
      perturbation. The fiducial model has three massless neutrinos,
      one of which is given a mass of 0.5 eV. Scales corresponding to
      the Jeans length at equality ($a=a_{\rm EQ}$) and today
      ($a=a_0$), as well as the Hubble horizon today are labeled.
\label{fig:nueffect}}
\end{center}
\end{figure}

The above discussion helps us understand the changes to the dark
matter perturbation evolution seen in Fig. \ref{fig:nueffect}. On
the smallest scales in Fig. \ref{fig:nueffect}, there is a uniform
suppression because neutrinos never cluster \cite{hu98}. On scales
larger than the Jeans length today, ${\cal O}({\rm Mpc})$ for 
$m_\nu={\cal O}({\rm eV})$, neutrinos do cluster once
$k \la k_J(a)$. This accounts for the increase in $\delta_m$ from about 10
Mpc to about 100 Mpc. For scales much larger than the Jeans length at
matter-radiation equality, which is ${\cal O}(100 {\rm Mpc})$ for  
$m_\nu = {\cal O}({\rm eV})$, the neutrinos can always cluster after
horizon crossing. In this regime the suppression of the dark matter 
perturbation comes from the change in the overall  growth factor
because the addition of a massive  neutrino speeds up the  expansion
rate of the universe.  

The amplitude of dark matter perturbations determines the fluctuations
in the metric $\Phi$ which in turn feeds into the lensing deflection
angle as $d(\vec L)=\sqrt{L(L+1)}\phi(\vec L)$.
$\phi(\vec L)$ is the fourier transform on the sky of 
$\phi(\vec \theta) = 2\int d\chi \Phi(\vec \theta \chi, \chi) 
(\chi_{\rm rec} - \chi)/(\chi\chi_{\rm rec}) $ where the integration
is along the unperturbed geodesic of the photon. We will denote the
three dimensional power spectrum of the metric perturbation $\Phi$ by
$P_\Phi(k)$. 

The lower panels in Figs. \ref{fig:TE} and \ref{fig:cldd} show the
differences in the power spectra when one of the three neutrinos is
given a mass of 0.1 eV. The error boxes are those for CMBpol
(described below; see Table I). Note that the signature of a 0.1 eV
neutrino in the angular power spectra, in the absence of lensing, is
at the 0.1\% level as shown in Fig. 1.

\newcommand{\bbnonu}{$\beta\beta 0 \nu$}
{\parindent0pt \bf Limits on Neutrino Mass.}
Presently, the most stringent laboratory upper bound on neutrino mass
comes from tritium beta decay end-point experiments \cite{tritium}
which limit the electron neutrino mass to $\la 2$ eV. 
Several proposed experiments plan to reduce this limit by one to two
orders of magnitude by searching for neutrino-less double beta decay
(\bbnonu) \cite{zdesenko03}.  A Dirac mass would elude this search,
but theoretical prejudice favors (and the see-saw mechanism requires)
Majorana masses. 
The theoretical uncertainties associated with \bbnonu \ experiments
are large \cite{vergados02} but this is the only way to ascertain if
the neutrino is a Majorana particle. 
Like the CMB and galaxy shear observations, these future \bbnonu \
experiments will be extremely challenging. 
We note that galaxy shear observations have the potential to be
competitive with the CMB and \bbnonu \ experiments in putting limits on
the neutrino mass \cite{hu99,abazajian02b}.

\begin{figure}[htbp]
  \begin{center}
    \plotone{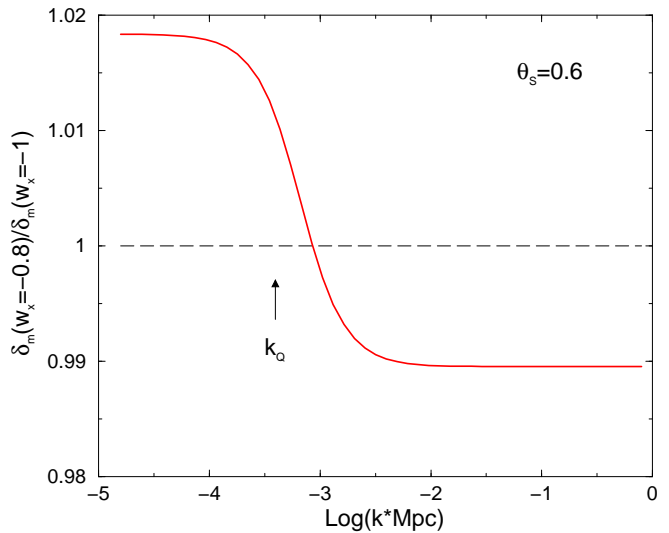}
    \caption{Effect on the dark matter perturbation of a change in the
      dark energy equation of state from $w_X=-1$ to $w_X=-0.8$.
\label{fig:weffect}}
\end{center}
\end{figure}

{\parindent0pt \bf Lensing and Dark Energy.}
Dark energy affects lensing in two distinct ways. First, the presence
of dark energy implies faster expansion and hence a decrease in the
overall growth rate. Second, dark energy can cluster appreciably on
scales $k \ga k_Q \equiv 2V_{,QQ}^{-1/2}$ \cite{ma99} where we have
modeled dark energy as a scalar field $Q$ with effective mass
$V_{,QQ}^{1/2}$ which is ${\cal O}(H_0)$. The clustering of dark
energy boosts the metric perturbations and hence lensing. However, if
the dark energy equation of state $w_X$ ($p_X=w_X \rho_X$ with $X$
denoting the dark energy component) is close to -1, then the
perturbations in the dark energy density are not important. The effect
of a change in $w_X$ is shown in  Fig. \ref{fig:weffect}. 

The $l$-dependence of $\partial \ln \cldd / \partial w_X$ is 
different from that of $\partial \ln \cldd / \partial m_\nu$ (see
Fig. \ref{fig:cldd}).  As mentioned, the primary effect is due to a
suppression of the overall growth factor and hence the effect of $w_X$ 
on $\cldd$ is not strongly $l$-dependent.
Note that the effect of $w_X$ will be more pronounced for larger
values of $w_X$ due to two reasons. One, dark energy starts to
dominate earlier (which implies larger uniform  suppression) and two,
perturbations in dark energy on large scales are enhanced for large
$w_X$.   

The CMB lensing window function is fairly broad in redshift--space. A
downside of this is the fact that CMB lensing will never be
competitive with SNIa observations (and possibly cosmic shear) as far
as measuring $w_X$ is concerned. However, the virtue of CMB lensing
is that it would be a robust alternative probe of the acceleration of
the universe and that the physics is different from that of SNIa since
lensing is sensitive to both the growth of perturbations and 
distances. The sensitivity to a broad range of redshifts also 
implies that CMB lensing is a unique probe of dark energy (more
generally clustering) at $z \ga 2$. We also note that if $w_X$ is
demonstrably different from -1, then dark energy must cluster on 1000
Mpc scales. The clustering properties of dark energy, say
parameterized in terms of it sound speed, might then be measurable. 

{\parindent0pt \bf Reionization, Optical Depth and Primordial Power
  Spectrum.} 
Reionization of the universe happened sometime between the redshifts
of 30 and 6.3. The lower bound comes from the observation of the
Gunn-Peterson trough in a $z=6.3$ quasar \cite{becker01}. The
upper bound comes from the upper limit to the value of the optical
depth to Thomson scattering, denoted by $\tau$, for photons from WMAP
TE two-point function measurements \cite{kogut03}. 

The CMB photons scatter off of the energetic free electrons produced
during reionization of the universe and this leaves an imprint on
their anisotropy which allows us to study the end of dark ages (the era
before the first stars turned on). A measurement of the optical depth
also allows us to determine the amplitude of the primordial power
spectrum $P_\Phi^i$ better. The overall power of the primary CMB
anisotropy on scales $l\ga 100$ is set by the combination 
$P_\Phi^i \exp(-\tau)$. It is this combination that present CMB
observations constrain well. Any information about $\tau$ therefore
breaks this $\tau$--$P_\Phi^i$ degeneracy and allows for a determination
of the primordial power spectrum amplitude. 

The scattering of CMB photons after reionization changes the primary
anisotropy at large angles or low $\ell$ corresponding to the horizon
during reionization \cite{zaldarriaga97b}. This change is not evident
in the TT power spectrum because of the dominant Sachs-Wolfe effect at
low $\ell$. However, there is no analog of the Sachs-Wolfe effect for
polarization and hence any new anisotropy produced at the
``reionization surface''  should be measurable through the TE and EE
power spectrum at low $\ell$. This bump in the power at low $\ell$
has been recently measured by the WMAP team \cite{kogut03} which
allowed them to ascertain that the universe was first reionized at 
$z \sim 20$.

Planck should measure this low $\ell$ signal much better. A
theoretical limiting factor to how well $\tau$ can be measured is our
ignorance about the reionization history \cite{kaplinghat02a,hu03} 
(i.e., evolution of the ionized hydrogen fraction between redshifts of
6 and 30). It was recently shown \cite{holder03}  that including
this ignorance implies that the 1-$\sigma$ error on $\tau$ from Planck
should be about 0.005 which implies that the amplitude of the
primordial power spectrum will be measured to about 1\%. 

{\parindent0pt \bf Lensing and Primordial Power Spectrum.}
Another way to break the $\tau$-$P_\Phi^i$ degeneracy is through the
small scale lensing signal \cite{hu02a}. Note that due to lensing the
power spectra do not scale as $P_\Phi^i$ on small scales. If lensing
were the dominant contribution to the two point functions, then
measured two point functions would scale as $(P_\Phi^i)^2$. In
addition, on small scales, one has to include the effect of non-linear
evolution of the density perturbations which have more complicated
dependence on  $P_\Phi^i$. Thus precision measurements of the small
scale CMB power can in fact measure $P_\Phi^i$ and thus provide an
indirect measurement of the optical depth. This effect would be
important if, for example, the low $\ell$ reionization bump cannot be
measured well.  

Measuring the lensing signal requires high angular
resolution. Along with the lensing signal we can also estimate the
primary CMB signal out to small scales. This is very useful for 
measuring the scale-dependence of the primordial power spectrum. In
particular if we take the primordial power spectrum to be 
$k^3P_\Phi^i(k) = k_f^3P_\Phi^i(k_f)(k/k_f)^{n_S -1+n_S'\ln(k/k_f)}$
with $k_f = 0.05/ {\rm Mpc}$, then future CMB experiments can provide
a precision measurement of $n_S$ and $n_S'$ (derivative of $n_S$ with
respect to $\ln(k)$). In the context of inflationary models we expect 
$n_S' \sim (n_S-1)^2$. An experiment like CMBpol will be able to test
this relation if $|n_S-1| \ga 5\%$. 

{\parindent0pt \bf Lensing, Reionization and Gravity Waves from 
Inflation.}    
The primary motivation for a full-sky mission after Planck, see CMBpol
in Table I, is the detection of the B mode due to gravity waves
produced in inflation.  The amplitude of this signal would directly
give us the energy density during inflation. At small scales this
primordial (tensor) B mode signal will be overwhelmed by the scalar B
mode signal unless the ratio of the primordial tensor to scalar power
spectrum amplitudes is larger than about 1\%. The scalar B mode signal
arises because (as we discussed before) lensing converts some E mode
polarization to B modes. Following the calculation 
in \cite{knox02} (see also \cite{kesden02}) we find a $3\sigma$
detection is possible for CMBpol if the fourth root of the energy
density during inflation is greater than $E_{\rm min} = 2 \times
10^{15}$ GeV which is an order of magnitude smaller than the GUT
scale. We note that $E_{\rm min}^4 \propto 1/\tau$, approximately, for  
$0.05 < \tau < 0.2$ and we have assumed $\tau=0.1$.  This scaling with
$\tau$ suggests that the largest angular scales, where the
reionization feature in the B mode appears, are important and
therefore a full-sky experiment is necessary to achieve this
sensitivity level. 

{\parindent0pt \bf Recombination, Helium Mass Fraction and BBN.}
The free electron fraction during recombination depends on the helium
mass fraction $Y_P$. Thus a change in the helium mass fraction changes
the visibility function at the recombination surface and hence 
the primary CMB power spectra. This implies that by measuring the
acoustic peaks out to $\ell \sim 1000$ well, one can determine
the helium mass fraction. In order to reach that conclusion, we have
to assume that the standard recombination physics is correct which
would not be the case, for example, if the fine structure constant
was different at the recombination surface \cite{kaplinghat99}.

Thus, along with the baryon density, future CMB experiments will
be able to determine $Y_P$ to high precision. This will facilitate 
precision consistency tests with Big Bang Nucleosynthesis (BBN)
predictions. It will also be useful in  constraining non-standard
BBN. For example, determining baryon density and helium mass fraction to
high precision allows strong constraints to be put on the number of 
relativistic species N (or equivalently the expansion rate) during
BBN. If $\sigma(Y_P)$ is small, then  $\sigma(N)=\sigma(Y_P)/0.013$,
which for CMBpol works out to $\sigma(N)=0.2$. Constraints on $N$ have
important repercussions for neutrino mixing in the early universe, and
hence on neutrino mass models \cite{abazajian02}. 

\begin{table}
\begin{center}
\begin{tabular}{cccccccc}
Experiment & $l_{\rm max}^{\rm UL}$ & $l_{\rm max}^{\rm T}$ &
$l_{\rm max}^{\rm E,B}$ & $\nu$ (GHz) & $\theta_b$ & $\Delta_T$ &
$\Delta_P$\\  
\tableskip\hline\tableskip
Planck   &2000   &2000  & 2500 &  100 & 9.2' & 5.5 & $\infty$ \\
          &    &    &  & 143 & 7.1' & 6  & 11 \\
          &    &    &  & 217 & 5.0' & 13 & 27 \\
\tableskip\hline
SPTpol ($f_{\rm sky} = 0.1$)&2000 & 2000 & 2500 &  217 & 0.9' & 12 &
17 \\ 
\tableskip\hline
CMBpol      &2000  & 2000 &  2500 & 217 & 3.0' & 1  & 1.4 \\
\tableskip\hline
\end{tabular}
\end{center}
\caption{Experimental specifications. Unlensed spectra ($\clttu$,
  $\clteu$, $\cleeu$) limited to $l < l_{\rm max}^{\rm UL}$. $\phi$
  reconstruction using only $l<l_{\rm max}^{T,E,B}$ lensed spectra.
} 
\end{table}

\begin{tablehere}
\begin{table*}[hbt]\small
\caption{\label{table:bounds}}
\begin{center}
{\sc Error Forecasts}\\
\begin{tabular}{c|c|c|c|c|c|c|c|c|c|c}
\tableskip\hline\hline \tableskip Experiment & $m_\nu$ (eV) & $w_x$ &
$\ln P_\Phi^i$ & $n_S$ & $n_S'$ & $\theta_s$ (deg) & $\tau$ & $\ln
\omega_m$ & $\ln \omega_b$ & $Y_P$ \\
\tableskip\hline\tableskip
Planck   &  0.15&  0.31&  0.017&  0.0071&  0.0032&  0.002&  0.0088&
0.0066&  0.0075& 0.012\\ 
SPTpol   &  0.18&  0.49&  0.018&  0.01&  0.006&  0.0026&  0.0088&
0.0087&  0.01& 0.017\\ 
CMBpol  &  0.044&  0.18&  0.017&  0.0029&  0.0017&  0.00064&  0.0085&
0.0022&  0.0028 &  0.0048\\
\tableskip\hline
\end{tabular}\\[12pt]
\begin{minipage}{5.2in}
Notes.---%
Standard deviations expected from Planck, SPTpol and CMBpol.
\end{minipage}
\end{center}
\end{table*}
\end{tablehere}

{\parindent0pt \bf Error Forecasting.}
The experiments we consider are Planck \cite{planck}, a
mission concept for a future full-sky mission called CMBpol
\footnote{CMBpol: http://spacescience.nasa.gov/missions/concepts.htm.},  
and a polarized bolometer array on the South Pole Telescope 
\footnote{SPT:  http://astro.uchicago.edu/spt/} we will call SPTpol.  
Their specifications are given in Table I.  
We assume that other frequency channels of Planck and CMBpol (not
shown in Table I) will clean out non-CMB sources of radiation
perfectly.  Detailed studies have shown foreground degradation of the
results expected from Planck to be mild 
\cite{knox99,tegmark00b,bouchet99}. We have restricted the range of
the power spectra keeping in mind that at small scales
foregrounds are likely to be an issue. Certainly at $l \ga 3000$ once
can expect emission from dusty galaxies to be a significant
contaminant. Other foregrounds like those due to kinetic SZ effect
from patchy reionization will also be severe at $l \ga 3000$, and may
even dominate the lensing signal \cite{knox98,santos03}. However,
these effects are less severe in polarization. We note that the
results for CMBpol are not significantly affected if we restrict
temperature data to $l < 1000$.   

The power spectra we include in our analysis are 
$\clttu$, $\clteu$, $\cleeu$ (unlensed), and $\cldd$. 
We do not use the lensed power spectra to avoid the complication of
the correlation in their errors between different $\ell$ values and
with the error in $\cldd$. Using the lensed spectra and neglecting
these correlations can lead to overly optimistic 
forecasts \cite{hu02a}. 

The error forecasting method \cite{jungman96,bond97} is based on the
fisher matrix formalism. Care must be exercised in calculating the
derivatives with respect to parameters like $w_X$ and $m_\nu$ since
their effects are small, and predominantly through lensing, once their
effect on the angular diameter distance is taken out. In particular,
the angle subtended by the sound horizon at the recombination surface,
$\theta_s$, must be included as one of the parameters since it will be
measured with exquisite precision. 
We take our parameter set to be ${\cal P} = \{\omega_m, \omega_b,
\omega_\nu, \theta_s, w_X,z_{\rm ri},k^3P_\Phi^i(k_f),n_s,n_s',Y_P\}$. 
The first three of these are the densities today (in units 
of $1.88\times 10^{-29}{\rm g}/{\rm cm}^3$) of cold dark matter plus
baryons, baryons and massive neutrinos. Note that for one massive
neutrino,  $\omega_\nu = m_\nu / 94.2 {\rm eV}$. 
The Thompson scattering optical depth for CMB photons, $\tau$, is
parameterized by the redshift of reionization $z_{\rm ri}$.  
We Taylor expand about 
${\cal P}=\{0.146,0.021,0,0.6,-1,6.3,6.4\times
10^{-11},1,0,0.24\}$. See \cite{kaplinghat03c} for more details on the
error forecasting method.

{\parindent0pt \bf Results.} 
The results are summarized in Table II and we list the salient features
here. Planck has the potential to measure the amplitude of the scalar
primordial power spectrum to about 1\% and inform us about the end of
dark ages by using the boost in power at low multipoles in the
CMB polarization due to reionization. For all entries in Table II we
have assumed this as prior knowledge and not included the low
multipoles in the analysis.   

Gravitational lensing of the CMB is clearly a promising probe of 
the growth of structure and the fundamental physics that affects
it. CMBpol can provide evidence for the acceleration of the universe,
independent of SNIa observations.  It can throw light on 
inflationary dynamics by testing for consistency with the relation
$n_S' \sim (n_S-1)^2$. The most striking result in Table II is the
promise that cosmology can provide us with an absolute determination
of neutrino mass to about 0.04 eV. We can be optimistic that a future
all-sky polarized CMB mission aimed at detecting gravitational waves
is likely to succeed in determining neutrino mass as well. 

{\parindent0pt \bf Acknowledgments.}
This article is based on work done in collaboration with Lloyd Knox
and Yong-Seon Song, and I would like to thank both my collaborators. I 
would also like to thank Wayne Hu for helpful discussions.

\end{document}